\documentclass[english]{article}
\usepackage[T1]{fontenc}
\usepackage[latin1]{inputenc}
\usepackage{a4wide}
\usepackage{babel}
\makeatletter

\providecommand{\LyX}{L\kern-.1667em\lower.25em\hbox{Y}\kern-.125emX\@}

\makeatother
\begin{document}
{\large \hfill{}}\textbf{\large Is it possible to clone using an}
\textbf{\emph{\large arbitrary}} \textbf{\large blank
state?\hfill{} }{\large \par}

\hfill{}Anirban Roy\( ^{a} \)\footnote{%
res9708@isical.ac.in
}, Aditi Sen(De)\( ^{b} \)\footnote{%
aditisendein@yahoo.co.in
} and Ujjwal Sen\( ^{b} \)\footnote{%
ujjwalsen@yahoo.co.in }\hfill{}

{\footnotesize \hfill{}\( ^{a} \)}\emph{\footnotesize Physics and
Applied Mathematics Unit, Indian Statistical Institute, 203 BT
Road, Kolkata 700035, India}{\footnotesize \hfill{}}
{\footnotesize \par{}}{\footnotesize \par}

{\footnotesize \hfill{}\( ^{b} \)}\emph{\footnotesize Department
of Physics, Bose Institute, 93/1 APC Road, Kolkata 700009,
India}{\footnotesize \hfill{}} {\footnotesize
\par{}}{\footnotesize \par}

\begin{abstract}
We show that in a cloning process, whether deterministic inexact
or probabilistic exact, one can take an arbitrary blank state
while still using a fixed cloning machine.
\end{abstract}
Quantum information cannot be cloned. There cannot exist a machine
which can produce two (or more) exact copies of an arbitrary state
in a deterministic manner \cite{1}. Replication is not allowed
even when the input state is taken at random from a given set of
two non-orthogonal states \cite{2}. It has been further shown that
probabilistic exact cloning is not possible if the input state is
from a given linearly dependent set \cite{3}.

Although exact cloning is not possible, one can approximately
clone an arbitrary input state \cite{4,5,6,7}. In this scheme a
\emph{fixed} state is taken as the blank state, depending on which
and the initial machine state, the cloning operation is
constructed. We extend this operation such that any
\emph{arbitrary} (pure or mixed) state, taken as the blank copy,
can do the job.

The optimal universal \( 1\rightarrow 2 \) inexact qubit cloner of
Bru\ss \emph{et al.}\cite{5} \emph{}takes an arbitrary qubit \(
\left| \psi \right\rangle \left\langle \psi \right|
=\frac{1}{2}(I+\overrightarrow{s}.\overrightarrow{\sigma }) \)
along with a fixed blank state \( \left| b\right\rangle  \) and a
machine state \( \left| M\right\rangle  \) as input. An entangled
state of the three qubits is produced as the output such that the
reduced density matrices of the first two qubits are two similar
approximate copies \( \rho =\frac{1}{2}(I+\eta
\overrightarrow{s}.\overrightarrow{\sigma }) \) of \( \left| \psi
\right\rangle \left\langle \psi \right|  \) with \( \eta
=\frac{2}{3} \). The unitary operator realizing this process is
defined on the combined Hilbert space of the input qubit, the
blank qubit and machine by\[ U^{\prime }\left| 0\right\rangle
\left| b\right\rangle \left| M\right\rangle
=\sqrt{\frac{2}{3}}\left| 00\right\rangle \left| m\right\rangle
+\sqrt{\frac{1}{6}}(\left| 01\right\rangle +\left| 10\right\rangle
)\left| m_{\perp }\right\rangle \]
\begin{equation}
\label{1} U^{\prime }\left| 1\right\rangle \left| b\right\rangle
\left| M\right\rangle =\sqrt{\frac{2}{3}}\left| 11\right\rangle
\left| m_{\perp }\right\rangle +\sqrt{\frac{1}{6}}(\left|
01\right\rangle +\left| 10\right\rangle )\left| m\right\rangle
\end{equation}
where \( \left| b\right\rangle  \) is a \emph{fixed} blank state
(in a two-dimensional Hilbert space), \( \left| M\right\rangle  \)
is the initial state of the machine, \( \left| m\right\rangle  \)
and \( \left| m_{\perp }\right\rangle  \) being two mutually
orthonormal states of the machine Hilbert space. The two clones
are to surface at the first and second qubits. Note that the
machine has turned out to be a qubit.

As it stands, the unitary operator \( U^{\prime } \) depends on
the blank state \( \left| b\right\rangle  \) and the machine state
\( \left| M\right\rangle  \). And the quality of the clones could
be badly affected if \( \left| b\right\rangle  \) gets changed to
an unknown state, say by some environment-induced decoherence. We
show that by suitably constraining the unitary operator it is
possible to keep the clones intact, even in this changed scenario.
After dealing with the \( 1\rightarrow 2 \) qubit cloner, we show
that the same is true for the most general cloner, the \(
N\rightarrow M \) \emph{qudit} (elements of a \( d \)-dimensional
Hilbert space) cloner. We carry over these considerations to the
case of probabilistic exact cloning.

Let us suppose that for the \( 1\rightarrow 2 \) qubit cloner, the
machine state \( \left| M\right\rangle  \) belongs to a
four-dimensional Hilbert space. And let the unitary operator \( U
\) be defined on the combined Hilbert space of the input qubit,
blank qubit and the four-dimensional Hilbert space of the machine
by\[ U\left| 0\right\rangle \left| b\right\rangle \left|
M\right\rangle =\sqrt{\frac{2}{3}}\left| 00\right\rangle \left|
M_{0}\right\rangle +\sqrt{\frac{1}{6}}(\left| 01\right\rangle
+\left| 10\right\rangle )\left| M_{1}\right\rangle \]
\[
U\left| 1\right\rangle \left| b\right\rangle \left| M\right\rangle
=\sqrt{\frac{2}{3}}\left| 11\right\rangle \left|
M_{1}\right\rangle +\sqrt{\frac{1}{6}}(\left| 01\right\rangle
+\left| 10\right\rangle )\left| M_{0}\right\rangle \]
\[
U\left| 0\right\rangle \left| b_{\perp }\right\rangle \left|
M\right\rangle =\sqrt{\frac{2}{3}}\left| 00\right\rangle \left|
M_{2}\right\rangle +\sqrt{\frac{1}{6}}(\left| 01\right\rangle
+\left| 10\right\rangle )\left| M_{3}\right\rangle \]
\[
U\left| 1\right\rangle \left| b_{\perp }\right\rangle \left|
M\right\rangle =\sqrt{\frac{2}{3}}\left| 11\right\rangle \left|
M_{3}\right\rangle +\sqrt{\frac{1}{6}}(\left| 01\right\rangle
+\left| 10\right\rangle )\left| M_{2}\right\rangle \] where \(
\left\langle M_{i}\right| \left. M_{j}\right\rangle =\delta _{ij}
\) (\( i,\, j=0,1,2,3) \) and \( \left\langle b\right| \left.
b_{\perp }\right\rangle =0 \).

Let \( \left| B\right\rangle =c\left| b\right\rangle +d\left|
b_{\perp }\right\rangle  \) be an arbitrary pure state of the
Hilbert space of the blank qubit. Then\[ U\left| 0\right\rangle
\left| B\right\rangle \left| M\right\rangle
=\sqrt{\frac{2}{3}}\left| 00\right\rangle \left| X\right\rangle
+\sqrt{\frac{1}{6}}(\left| 01\right\rangle +\left| 10\right\rangle
)\left| X^{\prime }\right\rangle \]
\begin{equation}
\label{2} U\left| 1\right\rangle \left| B\right\rangle \left|
M\right\rangle =\sqrt{\frac{2}{3}}\left| 11\right\rangle \left|
X^{\prime }\right\rangle +\sqrt{\frac{1}{6}}(\left|
01\right\rangle +\left| 10\right\rangle )\left| X\right\rangle
\end{equation}
where\[ \left| X\right\rangle =c\left| M_{0}\right\rangle +d\left|
M_{2}\right\rangle \]
 \[
\left| X^{\prime }\right\rangle =c\left| M_{1}\right\rangle
+d\left| M_{3}\right\rangle \] are orthogonal. This form is
exactly the same as in equation (1). Thus an arbitrary input qubit
\( \left| \psi \right\rangle  \) would be just as well cloned by
equation (2) as it would be through equation (1).

We now consider the most general cloning machine, the one that
produces \( M \) approximate copies of the input, when \( N(<M) \)
\( d \)-dimensional inputs are provided \cite{7}. The
corresponding unitary operator
need only be defined on the symmetric subspace\footnote{%
The symmetric subspace is defined by the linear span of the set of
all tensor product states \( \left| \psi \right\rangle \otimes
\left| \psi \right\rangle \otimes ...N \) times, \( \left| \psi
\right\rangle  \) being any \( d \)-dimensional state. } of the \(
d^{N} \)-dimensional Hilbert space of the \( N \) input qudits. It
is defined by \cite{7} \begin{equation} \label{3} U^{\prime
}_{NM}\left| \overrightarrow{n}\right\rangle \otimes \left|
R\right\rangle =\sum ^{M-N}_{\overrightarrow{j}=0}\alpha
_{\overrightarrow{n}\overrightarrow{j}}\left|
\overrightarrow{n}+\overrightarrow{j}\right\rangle \otimes \left|
M_{\overrightarrow{j}}\right\rangle
\end{equation}
where \( \overrightarrow{n}=(n_{1},\: n_{2},\: ....,\: n_{d}) \),
\( \left| \overrightarrow{n}\right\rangle  \) is a completely
symmetric and normalised state with \( n_{i} \) systems in \(
\left| i\right\rangle  \) with \( \sum ^{d}_{i=1}n_{i}=N \), \(
\overrightarrow{j}=(j_{1},\: j_{2},\: ....,\: j_{d}) \) with \(
\sum ^{d}_{k=1}j_{k}=M-N \), \( \left| R\right\rangle  \) denoting
the combined state of the \( M-N \) fixed pure \( d \)-dimensional
blank states \( \left| b_{d}\right\rangle  \) and the initial
state \( \left| M\right\rangle  \) of the cloning machine and \(
\left| M_{\overrightarrow{j}}\right\rangle  \) denoting the
orthonormal states of the cloning machine. And \[ \alpha
_{\overrightarrow{n}\overrightarrow{j}}=\sqrt{\frac{(M-N)!(N+d-1)!}{(M+d-1)!}}\sqrt{\prod
^{d}_{k=1}\frac{(n_{k}+j_{k})!}{n_{k}!j_{k}!}}\] There are \( s \)
equations required to define \( U^{\prime }_{NM} \), where \( s \)
is the dimension of the symmetric subspace \cite{key-8}. The
required dimension of the machine is \(
D=\frac{(M-N+d-1)!}{(M-N)!(d-1)!} \).

Now we proceed as we had done for the \( 1\rightarrow 2 \) qubit
cloner. If we want to allow an arbitrary pure state (possibly
entangled) of the \( d^{M-N} \)-dimensional Hilbert space of the
blank states to act as the new blank state and still produce the
same outputs, we have to use a \( (D\times d^{M-N}) \)-dimensional
machine. The new unitary operator \( U_{NM} \) satisfies, along
with the \( s \) equations in (3), \( (d^{M-N}-1)s \) more
equations corresponding to the \( d^{M-N}-1 \) more blank states
on which the new operator is to be defined. Then by linearity, for
an arbitrary pure blank state \( \left| B_{1}\right\rangle  \), we
would have\begin{equation} \label{4} U_{NM}\left|
\overrightarrow{n}\right\rangle \left| B\right\rangle \left|
M\right\rangle =\sum ^{M-N}_{\overrightarrow{j}=0}\alpha
_{\overrightarrow{n}\overrightarrow{j}}\left|
\overrightarrow{n}+\overrightarrow{j}\right\rangle \otimes \left|
X_{\overrightarrow{j}}\right\rangle
\end{equation}
where \( X_{\overrightarrow{j}} \) are orthonormal states of the
machine. This has the same form as eq. (3) and thus would be
equally efficient in producing the requisite approximate copies.

Similar considerations carry over to the case of probabilistic
exact cloning. Although we only consider the case qubits, the
considerations carry over to higher dimensions. In the case of
qubits, instead of the \( U_{1}^{\prime } \) defined by\cite{3}\[
U_{1}^{\prime }\left| \psi _{0}\right\rangle \left| b\right\rangle
\left| M\right\rangle =\sqrt{\gamma }\left| \psi _{0}\right\rangle
\left| \psi _{0}\right\rangle \left| m\right\rangle
+\sqrt{1-\gamma }\left| \Phi \right\rangle \]
\[
U_{1}^{\prime }\left| \psi _{1}\right\rangle \left| b\right\rangle
\left| M\right\rangle =\sqrt{\gamma }\left| \psi _{1}\right\rangle
\left| \psi _{1}\right\rangle \left| m\right\rangle
+\sqrt{1-\gamma }\left| \Phi \right\rangle \] with \( \gamma
=1/\left( 1+\left| \left\langle \psi _{0}\right| \left. \psi
_{1}\right\rangle \right| \right)  \) (\( \left| m\right\rangle
\) and \( \left| \Phi \right\rangle  \) are orthogonal), \( \left|
\psi _{0}\right\rangle  \), \( \left| \psi _{1}\right\rangle  \)
being two non-orthogonal states which are to be probabilistically
cloned, we define \( U_{1} \) as \[ U_{1}\left| \psi
_{0}\right\rangle \left| b\right\rangle \left| M\right\rangle
=\sqrt{\gamma }\left| \psi _{0}\right\rangle \left| \psi
_{0}\right\rangle \left| M_{0}\right\rangle +\sqrt{1-\gamma
}\left| \Phi \right\rangle \]
\[
U_{1}\left| \psi _{1}\right\rangle \left| b\right\rangle \left|
M\right\rangle =\sqrt{\gamma }\left| \psi _{1}\right\rangle \left|
\psi _{1}\right\rangle \left| M_{0}\right\rangle +\sqrt{1-\gamma
}\left| \Phi \right\rangle \]
\[
U_{1}\left| \psi _{0}\right\rangle \left| b_{\perp }\right\rangle
\left| M\right\rangle =\sqrt{\gamma }\left| \psi _{0}\right\rangle
\left| \psi _{0}\right\rangle \left| M_{1}\right\rangle
+\sqrt{1-\gamma }\left| \Phi ^{\prime }\right\rangle \]
\[
U_{1}\left| \psi _{1}\right\rangle \left| b_{\perp }\right\rangle
\left| M\right\rangle =\sqrt{\gamma }\left| \psi _{1}\right\rangle
\left| \psi _{1}\right\rangle \left| M_{1}\right\rangle
+\sqrt{1-\gamma }\left| \Phi ^{\prime }\right\rangle \] where \(
\left| M_{0}\right\rangle  \), \( \left| M_{1}\right\rangle  \),
\( \left| \Phi \right\rangle  \), \( \left| \Phi ^{\prime
}\right\rangle  \) are mutually orthogonal. Then \[ U_{1}\left|
\psi _{0}\right\rangle \left| B\right\rangle \left| M\right\rangle
=\sqrt{\gamma }\left| \psi _{0}\right\rangle \left| \psi
_{0}\right\rangle \left| m^{\prime }\right\rangle +\sqrt{1-\gamma
}\left| \Phi ^{\prime \prime }\right\rangle \]

\[
U_{1}\left| \psi _{1}\right\rangle \left| B\right\rangle \left|
M\right\rangle =\sqrt{\gamma }\left| \psi _{1}\right\rangle \left|
\psi _{1}\right\rangle \left| m^{\prime }\right\rangle
+\sqrt{1-\gamma }\left| \Phi ^{\prime \prime }\right\rangle \] for
an arbitrary blank qubit \( \left| B\right\rangle =c\left|
b\right\rangle +d\left| b_{\perp }\right\rangle  \) so that \(
\left| m^{\prime }\right\rangle =c\left| M_{0}\right\rangle
+d\left| M_{1}\right\rangle  \) and \( \left| \Phi ^{\prime \prime
}\right\rangle =c\left| \Phi \right\rangle +d\left| \Phi ^{\prime
}\right\rangle  \) are orthogonal states. Consequently, the
probabilistic cloning goes through with the same optimal
efficiency even if we use an arbitrary blank pure qubit.

As we have mentioned, the main motivation behind consideration of
an arbitrary blank state was decoherence. Decoherence, however,
usually produces \emph{mixed} states. But in our discussion we
have only considered pure states. We now show that the input blank
state could as well be an arbitrary mixed state. For definiteness,
let us consider only mixed qubits. Any such mixed state can be
written as \( \rho =a_{1}\left| a_{1}\right\rangle \left\langle
a_{1}\right| +a_{2}\left| a_{2}\right\rangle \left\langle
a_{2}\right|  \) where \( a_{1},\: a_{2}\geq 0 \), \(
a_{1}+a_{2}=1 \) and \( \left\langle a_{1}\right. \left|
a_{2}\right\rangle =0 \). Since \( \left| a_{1}\right\rangle  \)
and \( \left| a_{2}\right\rangle  \) are pure states, by linearity
it follows from (2) that the same unitary operator that allowed an
arbitrary pure blank state, would just as well clone even when the
blank state is a mixed qubit.

To summarize, we have shown that an unknown blank state can be
used for cloning, whether it is deterministic inexact or
probabilistic
exact.\\

We thank the anonymous referee for making useful suggestions to
revise our earlier manuscript. We acknowledge Guruprasad Kar,
Sibasish Ghosh, Debasis Sarkar, Pinaki Pal and Mridula Kanoria for
helpful discussions. A.S. and U.S. thanks Dipankar Home for
encouragement and U.S. acknowledges partial support by the Council
of Scientific and Industrial Research, Government of India, New
Delhi.


\begin{thebibliography}{1}
\bibitem{key-8}{\small The dimension \( s \) of the symmetric subspace of \( N \)
\( d \)-level systems is \( s=(N+d-1)!/N!(d-1)!
\)}
\end{thebibliography}
\end{document}